\documentclass[pre,showpacs,showkeys,amsmath,amssymb,twocolumn,a4paper]{revtex4}

\usepackage{graphicx}
\usepackage[latin1]{inputenc}

\begin{document}

\title{Diluted planar ferromagnets: nonlinear excitations on a non-simply connected manifold}

\author{ Fagner M. Paula} \author{Afranio R. Pereira}\thanks{Corresponding author : Afranio Rodrigues Pereira.\\
Tel.: 55-31-3899-2988;    fax: 55-31-3899-2483.} \email{apereira@ufv.br} \author{Lucas A. S. Mól}\email{lucasmol@ufv.br} \affiliation{Departamento de F\'isica, Universidade Federal de Vi\c cosa, Vi\c cosa,  36570-000, Minas Gerais, Brazil.}

\begin{abstract}
We study the behavior of magnetic vortices on a
two-dimensional support manifold being not simply connected. It is
done by considering the continuum approach of the XY-model on a
plane with two disks removed from it. We argue that an effective
attractive interaction between the two disks may exist due to the
presence of a vortex. The results can be applied to diluted planar
ferromagnets with easy-plane anisotropy, where the disks can be
seen as nonmagnetic impurities. Simulations are also used to test
the predictions of the continuum
limit.\end{abstract}

\pacs{75.10. Hk; 75.10.-b; 05.45.Yv; } \keywords{Impurities, Vortices, XY-model.}

\maketitle

Physics in two spatial dimensions has generated a lot of
surprises. In condensed matter systems these surprises are
specially interesting due to the possibility of technological
applications. The role of nonlinear excitations in the study of
low-dimensional, artificially structured materials is a very
exciting topic due to their observable effects on the physical
properties of a realizable condensed matter system.
Nonconventional supports (such as curved surfaces) for condensed
matter materials may induce a much richer physics because the
interplay between geometry and topology \cite{1,2,3,4}. Coulomb systems
living on a sphere \cite{5,6}, magnetic systems living on a cylinder or
a sphere \cite{1,2,3,4} etc have already been considered recently. Even
flat surfaces (the plane $R^{2}$) with pieces cut out from them
open up new, interesting avenues of investigation. For example, if
one or more sectors are excised from a single layer of graphite
and the remainder is joined seamlessly, a cone results \cite{7,8}. By
considering the symmetry of a graphite sheet and the Euler's
theorem, it can be shown that only five types of cone can be made
from a continuous sheet of graphite corresponding to the following
values of cone angles $\gamma=19.2^{o}, 38.9^{o}, 60.0^{o},
84.6^{o}, 112.9^{o}$. These are all synthesized forming carbon
nanocones \cite{8}. The conical geometry is also interesting because of
its connection with the problem of Einstein gravity in low
dimensions \cite{9}. Another interesting possibility is the plane
$R^{2}$ with a disk of radius $a$ cut out from it. Such a space is
said to be not simply connected since it has the property that
some closed curves drawn in it can not be continuously shrunk to a
point. In two-dimensional magnetic materials described by the
continuum limit, such a disk cut out from the plane has a simple
physical interpretation. It can be seen as a nonmagnetic impurity
present into the system \cite{10,11,12,13}. Then, from now, it is convenient
to use the language of magnetism.

A large variety of two-dimensional magnetic materials are
well described by the anisotropic Heisenberg models. Particularly,
many of these materials are modelled as a continuum of classical
spins with easy-plane symmetry. In these systems, vortices are
important excitations, responsible for interesting static and
dynamic properties \cite{14,15,16,17}. In this letter, we explore the effect
of the support manifold (the plane $R^{2}$) not simply connected
and focus on the vortex behavior. Specifically, we consider the
plane $R^{2}$ with two disks of radius $a$ cut out from it. In the
case of magnetism, the size $a$ is the lattice constant. We note,
however, that in experimental situations involving small scales
(e.g., a small hole with size of the order of lattice spacing $a$)
the applications of the continuum approximation (i.e., a long
wavelength theory) should be viewed with caution. Then, we also
compare some results of this theory with numerical simulations on
a discrete lattice. Our main motivation is to know how magnetic
vortices could be pinned in the system. Recent works \cite{10,11,12,13} have
shown that the topological excitations center is attracted and
pinned by a nonmagnetic impurity. However, a real magnetic
material contains many vacancies and then, the possibility of
pinned vortices not centered on vacancies but among them should be
investigated. Here, first we consider two spin vacancies and a
single vortex to calculate the equilibrium position for the vortex
center. The configurations of energy minima are then obtained. It
is shown that the effective potential experienced by a vortex due
to the presence of the two lattice defects attracts the vortex
center to either one of the vacancies center or the center of the
line joining the two vacancies. Second, we argue that the presence
of the magnetic vortex background may induce an effective
interaction between the holes that is attractive. This interaction
results in a geometrical frustration in the region between
vacancies.

The easy-plane ferromagnets are described by the
Hamiltonian
\begin{equation}\label{eq:eq1}
 H=-J\sum_{i,j}[S_{i}^{x}S_{j}^{x}+S_{i}^{y}S_{j}^{y}+\lambda S_{i}^{z}S_{j}^{z}],
\end{equation} where $J>0$ is the exchange constant, $0\leq\lambda<1$ is the
easy-plane anisotropy and $
\vec{S_{i}}=\{S_{i}^{x},S_{i}^{y},S_{i}^{z}\}$ is the classical
spin vector at site $i$. We focus on the case $\lambda=0$
(XY-model). The spin field can be parametrized by two scalar
fields $\phi$ and $ m=cos\theta$ (the azimuthal and polar angles)
as follows $ \vec{S}= \{\sqrt{1-m^{2}}cos\phi,
\sqrt{1-m^{2}}sin\phi,m\}$. The effective interaction among a
single vortex and the two spinless sites is discussed by using a
simple continuum XY-Hamiltonian on a support manifold
being not simply connected, written as \\
\begin{eqnarray}\label{eq:eq2}
H_{I}=\dfrac{J}{2}\int \left[
\dfrac{m^{2}(\vec{\nabla}\phi)^{2}}{1-m^{2}}+(1-m^{2})(\vec{\nabla}\phi)^{2}+\dfrac{4}{a^{2}}m^{2}\right]\times \nonumber \\ U_{1}(\vec{r})U_{2}(\vec{r})d^{2}r, \qquad \qquad \qquad
\end{eqnarray} \\
where the term in brackets is the Hamiltonian density $
\textsf{H}$ for the continuum version of pure Hamiltonian (1), the
function $U_{1}(\vec{r})$ defined as $U_{1}(\vec{r})=1 $ if $
\vert \vec{r}-\vec{r}_{1}\vert\ge a $ and $ U_{1}(\vec{r})=0 $ if
$ \vert \vec{r}-\vec{r}_{1}\vert < a $, represents the spin
vacancy (the first disk) centered at position $\vec{r}_{1}$ and
the function $U_{2}(\vec{r})$ defined as $U_{2}(\vec{r})=1 $ if $
\vert \vec{r}-\vec{r}_{2}\vert\ge a $ and $ U_{2}(\vec{r})=0 $ if
$ \vert \vec{r}-\vec{r}_{2}\vert < a $, represents another spin
vacancy (the second disk) centered at position $\vec{r}_{2}$. The
distance between vacancies is $p=\mid \vec{r}_{2}-\vec{r}_{1}\mid
$. We now make a very simple mathematical observation that follows
from the associative property of the terms present in expression
(2) and that justify the equivalency between a hole on the plane
and a vacancy in the Hamiltonian. Note that $\int
\textsf{H}[U_{1}(\vec{r})U_{2}(\vec{r})d^{2}r]= \int
[\textsf{H}U_{1}(\vec{r})U_{2}(\vec{r})]d^{2}r$ and then, the
problem of a magnetically coated plane not simply connected (left
side of the identity) is equivalent to the problem of a magnetic
plane simply connected but containing spin vacancies in the
Hamiltonian (right side). It only depends on the point of view:
the disks are removed either from the plane or from the
Hamiltonian density. With this in mind, we remember that
simulations show that the vortex structure is not deformed
appreciably due the presence of a hole \cite{13}. It will also be
assumed in our calculations.

The XY-model supports only planar vortices \cite{19}, which the
configuration is $m_{v}=0, \phi_{v}=\arctan[(y-y_{v})/(x-x_{v})]$
for the vortex center localized at $(x_{v},y_{v})$. The energy of
this vortex configuration, in the absence of nonmagnetic
impurities, is $ E_{v}= \pi J \ln(L/0.24a) $ \cite{19}, where $L$ is
the system size. Then, considering the vortex center localized at
origin (for simplicity), the effective interaction between the
vortex and the pair of impurities is given by $
V_{eff}(\vec{r}_{1},\vec{r}_{2})=E(\vec{r}_{1},\vec{r}_{2})-E_{v}
$, where $E(\vec{r}_{1},\vec{r}_{2})$ is the vortex energy in the
presence of the two impurities placed at $\vec{r}_{1}$ and
$\vec{r}_{2}$ respectively. Using Hamiltonian (2), the planar
vortex energy in the presence of two holes (nonmagnetic
impurities) is simply obtained leading to
\begin{eqnarray}\label{eq:eq3}
 E(\vec{r}_{1},\vec{r}_{2})= E_{v}+ \frac{\pi J}{2}\ln \left[1-\frac{a^{2}}{r_{1}^{2}+d^{2}}-\frac{a^{2}}{r_{2}^{2}+d^{2}}+ \right. \nonumber \\ \left. \frac{a^{4}}{(r_{1}^{2}+d^{2})(r_{2}^{2}+d^{2})}\right], \qquad \quad
\end{eqnarray} \\
where $d$ is a suitable constant of the order of the lattice
spacing $a$ introduced in Eq.(3) in order to avoid spurious
divergences of the vortex energy (in the continuum approach) when
the vortex center coincides with an impurity center. Thus, to find
this constant, we note that, in the case in which the centers of
the three defects (two defects in the lattice and one in the spin
field) are located at the same point (for example, at origin),
then, the three "bodies" problem reduces to a simpler problem of a
vortex on a single hole. In this case, $V_{eff}(\vec{0},\vec{0})$
obtained from Eq.(3), should have the value $-4.48J$ (see Ref.
\cite{13}) and hence, we get $d=1.1472a$. Although the limit that the
separation vanishes is reasonable in the continuum approach, in a
real discrete lattice two vacancies can not occupy the same point.
Therefore, these approximations can give us important insights
about the system with topological excitations and nonmagnetic
impurities.

The object now is to calculate the configurations of
minimum energy involving these three defects. It means that we
need to minimize the quantity
\begin{eqnarray}\label{eq:eq4}
F_{d}(r_{1},r_{2})=-\frac{1}{r_{1}^{2}+d^{2}}-\frac{1}{r_{2}^{2}+d^{2}}+\frac{a^{2}}{(r_{1}^{2}+d^{2})(r_{2}^{2}+d^{2})}.
\end{eqnarray} \\
To do this, we consider the geometry shown in Fig.(1). 
\begin{figure}
\includegraphics[height=2.5cm, keepaspectratio]{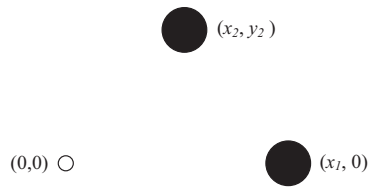}
\caption{A vortex (white circle) located at origin
$(0,0)$ near two holes (black circles) located at $
\vec{r}_{1}=(x_{1},0) $ and $ \vec{r}_{2}=(x_{2},y_{2})$. The
distance between the holes is
$p=\mid \vec{r}_{2}- \vec{r}_{1}\mid $.}
\end{figure} In this
figure, the vortex center is located at the origin and the
coordinates of the nonmagnetic impurities are $(x_{1},0)$ and
$(x_{2},y_{2})$. Here we have chosen $y_{1}=0$ (i.e., the hole 1
located along the x-axis) for effect of simplicity. Then, Eq.(4)
is rewritten as
\begin{eqnarray}\label{eq:eq5}
F_{d}(x_{1},x_{2},y_{2})=\frac{a^{2}-2d^{2}-x_{1}^{2}-x_{2}^{2}-y_{2}^{2}}{(x_{1}^{2}+d^{2})(x_{2}^{2}+y_{2}^{2}+d^{2})}.
\end{eqnarray} \\
Using the Lagrangian multipliers method, one has
$\vec{\nabla}F_{d}(x_{1},x_{2},y_{2})=-\kappa
\vec{\nabla}\varphi(x_{1},x_{2},y_{2}),$ with
$\varphi(x_{1},x_{2},y_{2})=p^{2}-(x_{2}-x_{1})^{2}-y_{2}^{2}=0$,
where $ \varphi$ is a constraint that keeps invariable the
distance $p$ between the two vacancies and $\kappa$ is a
parameter. Thus, we get the following system of equations to be
solved for $x_{1},x_{2}$ and $y_{2}$
\begin{eqnarray}\label{eq:eq6}
\left\{\begin{array}{cc}
x_{1}\frac{(a^{2}-d^{2}-x_{2}^{2}-y_{2}^{2})}{(x_{1}^{2}+d^{2})^{2}(x_{2}^{2}+y_{2}^{2}+d^{2})}-\kappa(x_{2}-x_{1})=0
& \\
\\
x_{2}\frac{(a^{2}-d^{2}-x_{1}^{2})}{(x_{1}^{2}+d^{2})(x_{2}^{2}+y_{2}^{2}+d^{2})^{2}}-\kappa(x_{2}-x_{1})=0
& \\
\\
y_{2}[\frac{a^{2}-d^{2}-x_{1}^{2}}{(x_{1}^{2}+d^{2})(x_{2}^{2}+y_{2}^{2}+d^{2})^{2}}+\kappa]=0
& \\
\\
p^{2}-(x_{2}-x_{1})^{2}-y_{2}^{2}=0
\end{array}\right.
\end{eqnarray}
We found two solutions for the above equations that satisfy our
requirement of extremum for $F_{d}$:
\begin{eqnarray}\label{eq:eq7}
(x_{1},x_{2},y_{2})=\left(\pm\frac{p}{2},\mp\frac{p}{2},0 \right)
\end{eqnarray} \\
or
\begin{eqnarray}\label{eq:eq8}
(x_{1},x_{2},y_{2})=(0,0,\pm p).
\end{eqnarray} \\
The identification of these solutions as energy minima will follow
from a consideration of the original equations. Substituting Eqs.
(7) and (8) into Eq. (3), one obtains the two energies $E_{1v}$
and $E_{2v}$ as a function of $p$ for these configurations,
respectively
\begin{eqnarray}\label{eq:eq9}
E_{1v}=E_{v}+ \frac{\pi J}{2}\ln \left(
1-\frac{a^{2}}{(p/2)^{2}+d^{2}}\right)^{2}
\end{eqnarray} \\
and
\begin{eqnarray}\label{eq:eq10}
E_{2v}= E_{v}+ \frac{\pi J}{2}\ln \left[
\left(1-\frac{a^{2}}{p^{2}+d^{2}}\right)\left(1-\frac{a^{2}}{d^{2}}\right)\right].
\end{eqnarray} \\
Hence, in general, any configuration of these three defects in
which the vortex center is found at the central point of the line
joining the two impurity centers, leads to the situation with
energy $E_{1v}$. By the other side, configurations of this system
in which the vortex center is found at the center of one of the
two impurities, produce the energy $E_{2v}$.

Now, it should be interesting to check the conditions of
stability of these two possibilities. It can be done by comparing
the energies $E_{1v}$ and $E_{2v}$ as a function of the impurities
separation $p$. Figure (2) shows the functions $\xi_{i}(p)$,
defined as $\xi_{i}(p)=E_{iv}-E_{v}$, $(i=1,2)$.
\begin{figure}
\includegraphics[height=4.5cm, keepaspectratio]{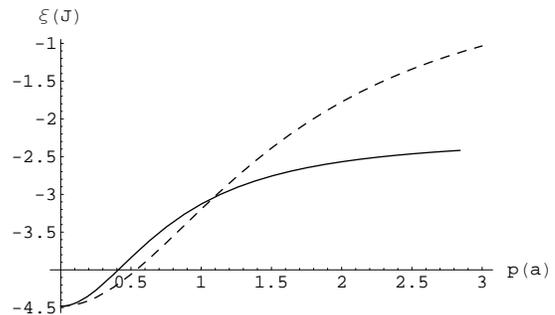}
\caption{Potentials $ \xi_{1}$ (dashed curve) and
$\xi_{2}$ (solid curve) versus $p$. For $p>1.087a$, $\xi_{2}< \xi_{1}$.}
\end{figure} Note that
$\xi_{1v}> \xi_{2v}$ for any $p > 1.087a$ and $\xi_{1v} <
\xi_{2v}$ for $0 < p < 1.087a$. It indicates that configuration 2
(with the vortex center coinciding with an impurity center) is
energetically favorable for an appreciable range of impurities
separation $p$. It is interesting to note that, in the case of the
two nearest neighbor vacancies (separated by the distance $a$),
the vortex center should occupy the center of the intersection of
holes (or the center of the large plaquette containing the two
neighbor spinless sites in a discrete lattice). Figure 2 also
shows that the potential $\xi_{2}$ is practically constant for
$p>2a$, and hence the force between the vortex-on-vacancy and the
other vacancy decays rapidly with the distance of separation. Note
also that, in this case, the energy necessary to remove a vortex
from a vacancy is $2.24J$, which is smaller than $4.48J$,
indicating that the presence of other impurities lower the biding
energy of the vortex-on-vacancy state. Extrapolating the present
results for quenched diluted layered magnetic materials with very
low impurity concentrations (in this case it is expect that the
vortex density is almost the same as the vacancy density), it is
conceivable that a magnetic vortex lattice (not necessarily
periodic, since the vortices distribution would be alike the
vacancies distribution) could be present into the system. Besides,
this vortex lattice could not be so rigid, because, as we have
seen, the impurity concentration must decrease the
vortex-on-vacancy binding energy, giving some mobility to
vortices. This picture would have important consequences to the
spin dynamics, which may be observed in experiments. For example,
the neutron scattering function $S_{\gamma\gamma}(\vec{q},\omega)$
may still exhibit a central peak (the cause of this peak is
believed to be the vortex translational motion \cite{15,16,17}), but its
shape, width and height should modify considerably.

Next, we investigate the nature of the interaction between
two static vacancies through the vortex background. In Ref.\cite{13},
it was shown that the energy of a vortex at origin in the presence
of only one impurity at distance $ \mid \vec{r}_{i} \mid $ away is
given by $ E_{v}+(\pi J/2)\ln[1-a^{2}/(r_{i}^{2}+b^{2})]$, where
$b=1.03a$. Hence, the energy cost of removing one spin (placed at
$\vec{r}_{i}$) from an infinite plane containing a vortex at
origin is $\epsilon_{1}( \vec{r}_{i})=-(\pi
J/2)\ln[1-a^{2}/(r_{i}^{2}+b^{2})]$. Of course, due to the
cylindrical symmetry, this energy does not depend on the direction
of $ \vec{r}_{i}$. From Eq. (3), it is easy to see that the energy
cost of removing two spins (located at sites $ \vec{r}_{1}$ and $
\vec{r}_{2}$) from the vortex background is
\begin{eqnarray}\label{eq:eq11}
 \epsilon_{2} ( \vec{r}_{1},\vec{r}_{2})= -\frac{\pi
 J}{2}\ln \left[1-\frac{a^{2}}{r_{1}^{2}+d^{2}}-\frac{a^{2}}{r_{2}^{2}+d^{2}}+\right. \nonumber \\ \left. \frac{a^{4}}{(r_{1}^{2}+d^{2})(r_{2}^{2}+d^{2})}\right].
\end{eqnarray} \\
Thus, there is an effective interaction potential between two
static holes through the vortex background given by $
\Delta=\epsilon_{2}( \vec{r}_{1},\vec{r}_{2})-\epsilon_{1}(
\vec{r}_{1})-\epsilon_{1}( \vec{r}_{2})$. This potential is, then,
expressed as
\begin{eqnarray}\label{eq:eq12}
 \Delta=\frac{\pi
 J}{2}\ln \left[\frac{1+a^{2}F_{b}(r_{1},r_{2})}{1+a^{2}F_{d}(r_{1},r_{2})}\right],
\end{eqnarray} \\
where $F_{b}(r_{1},r_{2})$ is also given by expression (4) but
with $d$ substituted by $b$. This effective interaction depends on
the distance of separation as well as the orientation of vacancies
in the vortex background. Since $b<d$, this energy is negative,
which means that the effective interaction between vacancies is
attractive. The problem can be summarized as follows: consider an
initial configuration of these three defects as shown, for
example, in Fig. (1). Supposing that $p>1.087a$, the spin
distribution (or the magnetic energy) will be then readjusted to
minimize the system energy, making the vortex center to move to
the nearest vacancy center. However, the spin distribution did not
reach its minimum in energy yet. The vortex-on-vacancy is still
attracted by the other vacancy, but now, neither the
vortex-on-vacancy state nor the other vacancy can dislocate on the
plane. In fact, in a rigid plane (lattice), the holes are fixed to
their original positions and can not move on the plane. As a
consequence, the system cannot reach its minimum in energy
anymore; there is a geometrical frustration, which stems from a
constraint (an invariable distance $p$ between the two
disconnected holes). Of course, the static-vacancy problem is an
important limiting model for understanding the interaction of
mobile holes through the magnetic background. If the vacancies
were mobile, an attractive interaction could result in holes
pairing around the magnetic plane. Probably, in an elastic
support, a variation in the geometry of the plane, compensating
for the constraint, may lead to a lowering of the energy. It is
expected that if the rigidity condition were relaxed, an elastic
Hamiltonian density introduced in the problem should stabilize the
plane against arbitrary deformations.

To test some results obtained by using the continuum
limit, we have also performed spin dynamical simulations on a
$L=20a$ square lattice to study the behavior of a single vortex
with its center initially located at the center of the system
$(0,0)$, in the presence of two spin vacancies located at
$(a/2,-3a/2)$ and $(-a/2,3a/2)$ respectively (see Fig. (3)).
\begin{figure}
\includegraphics[height=6cm, keepaspectratio]{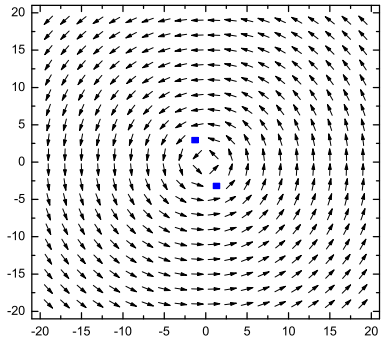}
\caption{Initial configuration with a vortex located at
origin and two vacancies at $(-a/2,3a/2)$ and $(a/2,-3a/2)$
respectively. The squares represent the spinless sites and the
distance between them is $p=3.16a$. The length scale is expressed in units of $a/2$.
Many other configurations were also studied confirming the analytical results.}
\end{figure} This
configuration starts with the vortex localized exactly at the
central point of the line joining the two vacancies. Note that the
analytical results indicate that it is an unstable situation,
since $p\approx 3.16a$, which is bigger than $1.087a$. Then, it is
expected that, in the simulations, the vortex center must move in
direction to one of the spinless sites. To see this we have
imposed diagonally antiperiodic boundary conditions \cite{20} $
\vec{S}_{L+1,y}=-\vec{S}_{1,L-y+1}$,
$\vec{S}_{0,y}=-\vec{S}_{L,L-y+1}$ and $
\vec{S}_{x,L+1}=\vec{S}_{L-x+1,1}$, $
\vec{S}_{x,0}=-\vec{S}_{L-x+1,L}$, for all $1\leq x,y\leq L$, in
order to keep the vortex structure stable. The discrete motion
equation for each spin is given by \cite{21} $d
\vec{S}_{i}/dt=\vec{S}_{i}\times \vec{h}$, where $
\vec{h}=-J\sum_{j}(S_{j}^{x}\hat{e}_{x}+S_{j}^{y}\hat{e}_{y})$ and
$\hat{e}_{x}$ and $\hat{e}_{y}$ are unit vectors in the $x$ and
$y$ directions, respectively. The motion equations were integrated
numerically forward in time using a fourth-order Runge-Kutta
scheme with a time step of $4 \times 10^{-4}J^{-1}$. We notice
that after $10^{4}$ time steps the position of the vortex center
always reaches one of the impurity centers (Fig. (4)).
\begin{figure}
\includegraphics[height=6cm, keepaspectratio]{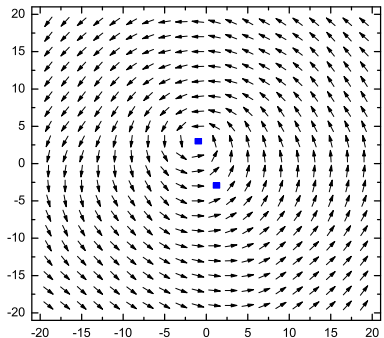}
\caption{Configuration after $7 \times 10^{4}$ time
steps. The vortex travelled from the central position between
vacancies to one of the vacancies center. It is in agreement with
analytical results.}\end{figure} On the
other hand, if the holes are separated by one lattice spacing, the
continuum theory predicts that the vortex center should stay on
the central point joining the holes. Our simulations also agree
with this result as it can be seen in figures 5 and 6. \begin{figure}
\includegraphics[height=6cm, keepaspectratio]{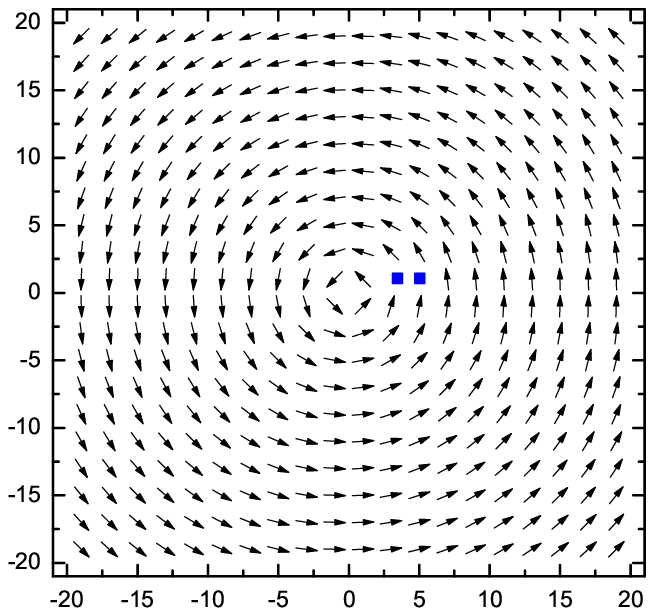}
\caption{Initial configuration with a vortex located at
origin and two vacancies at $(3a/2,a/2)$ and $(5a/2,a/2)$
respectively.
The distance between these two vacancies is $1a$.}\end{figure}
\begin{figure}
\includegraphics[height=6cm, keepaspectratio]{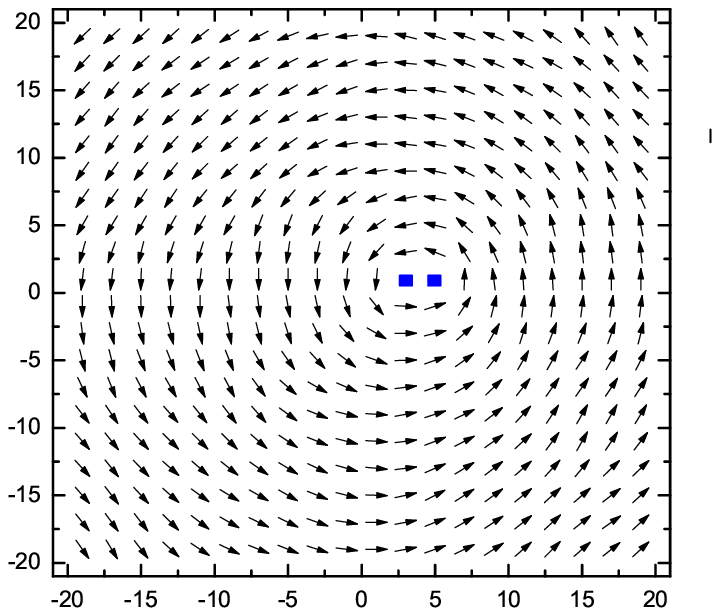}
\caption{Final state of Fig.(5). Now, the vortex center
stays in the central point of the line joining the two vacancies
in agreement with analytical results.}\end{figure} Figure (5)
shows an initial configuration with a vortex at origin and two
neighbor vacancies at $(3a/2,a/2)$ and $ (5a/2,a/2)$ respectively.
After $10^{4}$ time steps, the vortex center stays between the
vacancies (Fig. (6)). However, there is the possibility that this
last configuration may not be completely static, but the vortex
center seems to oscillate around the central point joining the
vacancies, affecting the spin dynamics. The analysis of this
possibility is out of the scope of this work
and will be investigated in a future paper.

In summary, we have studied the problem of the magnetic
vortex behavior in a non-simply connected manifold. The results
can be applied to magnetic materials, including antiferromagnetic
systems. The flat plane considered here was constructed by cutting
two holes, which can be interpreted as nonmagnetic impurities. The
cut edges (which are the boundary surfaces for the infinite
two-dimensional plane) exert a profound influence on the vortex
center. There are two ways of pinning a vortex, in which the
stability depends on the distance $p$ between the holes. Besides,
we have shown that the vortex background induces an effective
interaction potential between two vacancies that is attractive. It
generates geometrical frustration. If the holes and the resulting
hole-vortex hybrids have metallic mobilities, then such an
effective interaction could be a pairing mechanism for
superconductivity in doped layered antiferromagnets. The overall
qualitative agreement between the analytical results using the
continuum approach on a plane with two holes and the simulations
on a discrete lattice with two missing spins is striking. Our
results may also have relevance to dilute two
dimensional Josephson junction arrays.

\begin{acknowledgments} We would like to thank Dr. S.A. Leonel and Dr. P.Z. Coura for
helpful discussions. This work was supported by CNPq (Brazilian
agency).\end{acknowledgments}


\thebibliography{21}
\bibitem{1} S. Villain-Guillot, R. Dandoloff, A. Saxena, Phys. Lett. A 188 (1994) 343.
\bibitem{2} R. Dandoloff, S. Villain-Guillot, A. Saxena, A. R. Bishop, Phys. Rev. Lett. 74 (1995) 813.
\bibitem{3} R. Dandoloff, A. Saxena, Eur. Phys. J. B 29 (2002) 265.
\bibitem{4} S. Villain-Guillot, R. Dandoloff, A. Saxena, A.R.
Bishop, Phys. Rev.B 52 (1995) 6712.
\bibitem{5} J.M. Caillol, J. Physique-Lettres 42,(1981) L245.
\bibitem{6} P.J. Forester, B. Jancovici, J. Stat. Phys. 84,
(1996) 337.
\bibitem{7} A. Krishnan, E. Dujardin, M.M.J. Treacy, J. Hugdahl,
S. Lynum, T.W. Ebbesen, Nature (London) 388 (1997) 451.
\bibitem{8} P.E. Lammert, V.H. Crespi, Phys. Rev. Lett. 85 (2000).
\bibitem{9} J.D. Brown, Lower Dimensional Gravity (World Scientific, New Jersey, 1988), and references therein.
\bibitem{10} A.R. Pereira, A.S.T. Pires, J. Magn. Magn. Mater. 257 (2003) 290.
\bibitem{11} L.A.S. Mól, A.R. Pereira, W.A. Moura-Melo, Phys. Rev. B 67 (2003) 132403.
\bibitem{12} A.R. Pereira, Phys. Lett. A 314 (2003) 102.
\bibitem{13} A.R. Pereira, L.A.S. Mól, S.A. Leonel, P.Z. Coura, B.V. Costa, Phys. Rev. B 68 (2003) 132409.
\bibitem{14} J.M. Kosterlitz, D.J. Thouless, J. Phys. C 6 (1973)
1181.
\bibitem{15} F.G. Mertens, A.R. Bishop, G.M. Wysin, C. Kawabata,
Phys. Rev. B 39 (1989) 591.
\bibitem{16} A.R. Pereira, A.S.T. Pires, M.E. Gouvea, B.V.
Costa, Z. Phys. B 39 (1992) 109.
\bibitem{17} M.E. Gouvea, G.M. Wysin, A.R. Bishop, F.G. Mertens,
Phys. Rev. B 39 (1989) 11840.
\bibitem{18} G.M. Wysin, Phys. Rev. B 49 (1994) 8780.
\bibitem{19} G.M. Wysin, Phys. Rev. B 54 (1996) 15156.
\bibitem{20} H. Kawamura, M. Kikuchi, Phys. Rev. B 47 (1993) 1134.
\bibitem{21} H.G. Evertz, D.P. Landau, Phys. Rev. B 54 (1996)
12302.
\end{document}